 \documentclass[preprint,onecolumn,amsmath,amssymb,aps,prb]{revtex4-1}

\usepackage{graphicx}
\usepackage{dcolumn}
\usepackage{bm}
 \usepackage{float} 

\usepackage{fullpage}
\usepackage{color}
\usepackage{subcaption}
\usepackage{multirow}
\usepackage{amsmath}
\usepackage{amssymb}
\usepackage{amsthm}
\usepackage[colorlinks   = true, urlcolor     = cyan, linkcolor    = blue, citecolor   = red]{hyperref}
\usepackage{bm}
\usepackage{pict2e}
\usepackage{epstopdf}
\usepackage{comment}
\usepackage{epsfig}
\usepackage{url}
\usepackage[ruled,vlined]{algorithm2e}

\usepackage{cancel}
\usepackage{amsmath}
\usepackage{amssymb}
\usepackage{etoolbox}
\usepackage{graphicx}
 \usepackage[english]{babel}
\usepackage{mathrsfs} 
\usepackage{enumitem}                                  

\usepackage{amsthm} 
\usepackage{siunitx}
\usepackage{ upgreek }
 \newcommand{\noop}[1]{}

\begin{document}


\title{Ab initio study on the electromechanical response of Janus transition metal dihalide nanotubes}

\author{Arpit Bhardwaj}
\affiliation{College of Engineering, Georgia Institute of Technology, Atlanta, GA 30332, USA}

\author{Phanish Suryanarayana}
\email{phanish.suryanarayana@ce.gatech.edu}
\affiliation{College of Engineering, Georgia Institute of Technology, Atlanta, GA 30332, USA}


\begin{abstract}
We study the electronic response of Janus transition metal dihalide (TMH) nanotubes to mechanical deformations using Kohn-Sham density functional theory. Specifically, considering twelve armchair and zigzag Janus TMH nanotubes that are expected to be stable from the phonon analysis of flat monolayer counterparts, we first compute their equilibrium diameters and then determine the variation in bandgap and effective mass of charge carriers with the application of tensile and torsional deformations. We find that the nanotubes undergo a linear and quadratic decrease in bandgap with tensile and shear strain, respectively. In addition, there is a continual increase and decrease in the effective mass of electrons and holes, respectively. We show that for a given strain, the change in bandgap for the armchair nanotubes can be correlated with the transition metal's in-plane $d$ orbital's contribution  to the projected density of states at the bottom of the conduction band. 
\end{abstract}


\maketitle
\section{Introduction}
Nanotubes are hollow cylindrical structures whose diameters are the nanometer scale, with  lengths that are  orders of magnitude larger. These quasi one-dimensional materials, more than thirty of which have been synthesized since the pioneering work for carbon nanotubes \cite{iijima1991helical}, are known to display enhanced and exotic mechanical, electronic, optical, and thermal properties/behavior relative to their bulk counterparts \cite{tenne2003advances, rao2003inorganic, serra2019overview}. Moreover, these properties differ based on the diameter/chirality  of the nanotubes \cite{ghosh2019symmetry, maiti2003bandgap, zhang2011helical, guo2005systematic, jia2006structure, yang2005electronic, mintmire1993properties, wang2017band, ding2002analytical, ouyang2002fundamental} and  can be further  controlled/tuned through external stimuli, such as addition of defects \cite{wang2019investigation, parashar2015switching, akdim2011bandgap}, electric field \cite{tien2005band, chen2004band}, magnetic field \cite{fedorov2010tuning}, temperature \cite{li2020unveiling, nath2003superconducting, tsuneta2003formation}, and mechanical deformations \cite{yang2000electronic, rochefort1999electrical, tombler2000reversible, yang1999band, kim2001electronic, kinoshita2010electronic, coutinho2009band, ghassemi2012field, qi2012strain, niquet2012effects}, making them particularly suited for technological applications.

Among the nanotubes that have been synthesized, a large fraction are from the transition metal dichalcogenide (TMD) group \cite{tenne2003advances, rao2003inorganic, serra2019overview}, which consists of materials of the form MX$_2$, where M and X denote the transition metal and chalcogen, respectively. Given the large number of transition metal (34 in number) and chalcogen (4 in number) combinations (136 in number) that are possible, it is likely that even more TMD nanotubes will be synthesized in the future. However, such nanotubes are generally multiwalled and have large diameters that are typically in the range of 10--40 nm \cite{tenne2003advances,  rao2003inorganic,  serra2019overview}, a consequence of the relatively large bending moduli of TMD monolayers \cite{kumar2020bending}. This limits the possibility of fascinating and novel properties that are associated with quantum confinement effects. Also, the large variations in the diameters makes systematic theoretical/computational studies extremely challenging.

The Janus TMD nanotube group \cite{varjovi2021janus}, which consists of materials of the form MXY, where X and Y are different chalcogens, overcome many of the limitations of the TMD group. In particular, given the asymmetry in the system, single-walled Janus nanotubes become energetically more favorable than their flat counterparts \cite{xiong2018spontaneous, bhardwaj2021elastic}, which significantly increases the likelihood of them being synthesized. Indeed, the WSSe nanotube has been recently synthesized \cite{sreedhara2022nanotubes}. Moreover, they have an energy minimizing diameter, typically in the range of 3--16 nm \cite{bhardwaj2021elastic}, which is much smaller than the corresponding values for TMD nanotubes, significantly increasing the likelihood of novel and exotic properties.  In particular, recent work on the electromechanical response of Janus TMD nanotubes has shown that they have potential applications in mechanical sensors and semiconductor switches \cite{bhardwaj2022strain}, similar to the case of TMD nanotubes \cite{bhardwaj2021strain}. It is therefore to be expected that at the very least, Janus TMD nanotubes also inherit the other  applications of their non-Janus counterparts, including nano-electromechanical (NEMS) devices \cite{yudilevichself, levi2015nanotube, divon2017torsional},  biosensors \cite{barua2017nanostructured}, superconductive materials \cite{nath2003superconducting, tsuneta2003formation}, and photodetectors \cite{unalan2008zno, zhang2012high, zhang2019enhanced}.

The Janus transition metal dihalide (TMH) nanotube group \cite{bolle2021structural}, which consists of materials of the form MXY, where X and Y are now different halogens,  are likely to possess fascinating and  exciting properties similar to those displayed in flat TMH monolayers and their Janus variants, e.g., FeCl\textsubscript{2} is piezoelectric ferromagnetic with the Curie temperature around room temperature \cite{saritas2022piezoelectric}, and  FeClBr and FeClF are ferrovalley materials based on their magnetic anisotropy \cite{li2021room, guo2022substantial}. Simultaneously, the nanotubes inherit the aforementioned advantageous features of being a Janus structure. However, the properties and behavior of these nanotubes has not been studied till now, providing the motivation for the current work. In particular, we study the electronic response of Janus TMH nanotubes to mechanical deformations using ab initio Kohn-Sham density functional theory (DFT) calculations. Specifically, considering twelve armchair and zigzag Janus TMH nanotubes that are expected to be stable, we first compute their equilibrium diameters and then determine the variation in bandgap and effective mass of charge carriers with the application of tensile and torsional deformations. Overall, we find that mechanical deformations represent a powerful tool for controlling the electronic properties of Janus TMH nanotubes.

The remainder of this article is organized as follows. In Section~\ref{Sec:Methods}, we list the Janus TMH nanotubes selected and detail the calculation of their electronic response to mechanical deformations. The results so obtained are presented and discussed in Section~\ref{Sec:Results}. Finally, we provide concluding remarks in Section~\ref{Sec:Conclusions}.


\section{Systems and methods} \label{Sec:Methods}

We consider zigzag and armchair variants of the following Janus TMH nanotubes: (i) M$=$\{Ti, Zr, Hf\} and X,Y$=$\{Cl, Br, I\}, with 2H-t symmetry \cite{nath2002nanotubes, bandura2014tis2}; and (ii) M$=$\{Fe\} and X,Y$=$\{Cl, Br, I\}, with 1T-o symmetry \cite{nath2002nanotubes, bandura2014tis2}, all having the lighter halogen on the inner side. These represent the set of all Janus TMH nanotubes that have been predicted to be thermodynamically stable from  first principles investigations \cite{bolle2021structural}.

\begin{figure}[h!]
        \centering
        \includegraphics[width=0.99\textwidth]{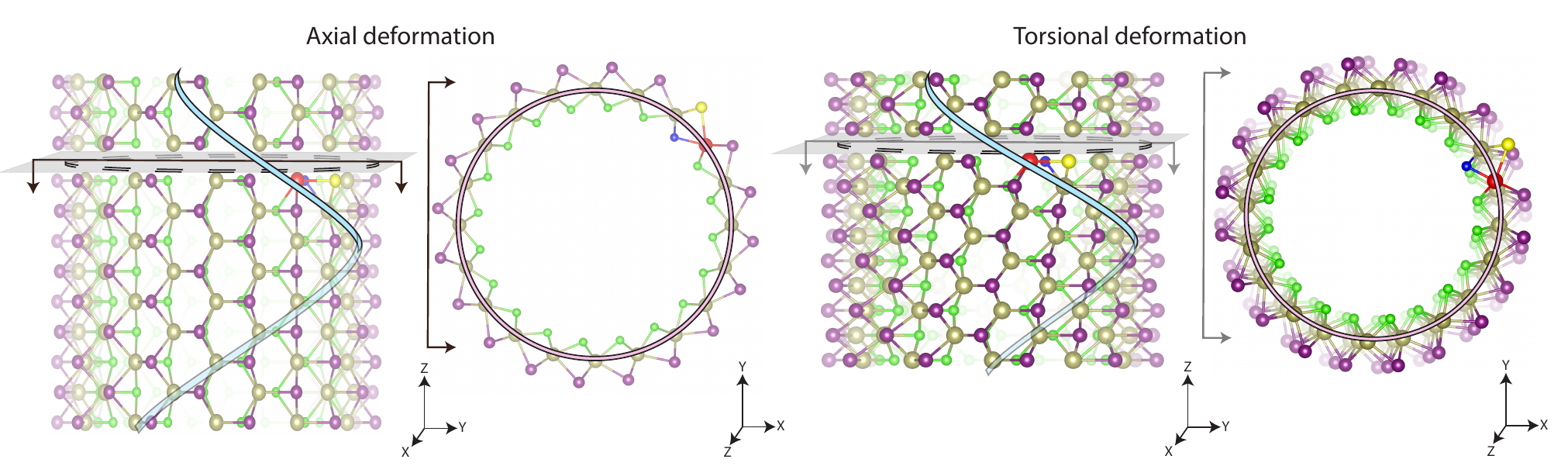}
        \caption{Illustration generated using VESTA  \cite{momma2008vesta} that  depicts the inherent cyclic and helical symmetry of an axially and torsionally deformed (10,10) 1T-o Janus TMH nanotube.  The entire nanotube can be generated using 3 atoms, e.g., metal and halogens colored red and blue/yellow, respectively, that lie within the cyclic+helical symmetry-adapted unit cell. This  symmetry is exploited while performing Kohn-Sham DFT calculations using the electronic structure code SPARC's Cyclix-DFT feature.}
      \label{fig:illustration}
    \end{figure}

We perform all nanotube simulations using the Cyclix-DFT \cite{sharma2021real} feature ---  well tested in various physical applications \cite{codony2021transversal, kumar2021flexoelectricity, sharma2021real, kumar2020bending, bhardwaj2021torsional, bhardwaj2021strain, bhardwaj2021elastic, bhardwaj2022strain, kumar2022bending} --- in the state-of-the-art real-space DFT code SPARC \cite{xu2021sparc, ghosh2017sparc1, ghosh2017sparc2}. In this formalism, as illustrated in figure~\ref{fig:illustration}, the cyclic and/or helical symmetry of the system is exploited to reduce all computations to a unit cell that contains only a small fraction of the atoms in the traditional periodic unit cell \cite{sharma2021real, ghosh2019symmetry, banerjee2016cyclic}, e.g.,  the periodic unit cell for a (45,45) HfClBr nanotube with diameter $\sim 9$ nm and an external twist of $6\times10\textsuperscript{-4}$ rad/Bohr has $169,155$ atoms,  whereas the cyclic+helical symmetry-adapted unit cell has only 3 atoms (one of each chemical element),  a number that remains unchanged by axial and/or torsional deformations. This symmetry-adaption provides tremendous computational savings, given that Kohn-Sham DFT computations scale cubically with system size.

In all simulations, we employ the Perdew–Burke–Ernzerhof (PBE) \cite{perdew1996generalized} exchange-correlation functional, and  scalar relativistic optimized norm-conserving Vanderbilt (ONCV) \cite{hamann2013optimized} pseudopotentials with nonlinear core correction from the SPMS collection \cite{spms}. The equilibrium configurations  for the flat monolayer counterparts so obtained (Supplementary Material) are in very good agreement with PBE results in literature  \cite{haastrup2018computational}, verifying the accuracy of the chosen pseudopotentials. Though PBE is known to generally  underpredict the bandgap \cite{haastrup2018computational},  it does provide good qualitative trends,  making it a common choice for DFT calculations and Janus transition metal nanotubes in particular \cite{wu2018tuning, tang2018janus, mikkelsen2021band, luo2019electronic, tao2018band, xie2021theoretical, wang2018mechanical}, motivating its selection here.  Even quantitatively,  sophisticated exchange-correlation functionals like hybrid  are not necessarily more accurate for Janus materials \cite{bhardwaj2022strain}.  Relativistic effects are neglected in all calculations,   since only minor band structure modifications  have been observed when spin-orbit coupling is incorporated for  TMH and Janus TMD monolayers  \cite{haastrup2018computational}.

We set the diameter of each nanotube to be that which minimizes the ground state  energy (Supplementary Material), the results for which are summarized in Table~\ref{Table:EqDia}.  All numerical parameters in Cyclix-DFT, including real-space grid spacing,  \emph{k}-point sampling for Brillouin zone, radial vacuum, and structural relaxation tolerances (both cell and atom) are chosen such that the lattice parameters and atomic positions are numerically converged to within 0.01 bohr, which translates to the energy being accurate to within $10^{-6}$ Ha/atom. We find that the equilibrium diameters so computed are in good agreement with Ref.~\cite{bolle2021structural}, the maximum difference of 0.6 nm occurring for the ZrClBr nanotube. The diameters follow the trend: MClI $<$ MBrI $<$ MClBr, which can be explained by the electronegativity difference in halogens, i.e., larger difference results in smaller equilibrium diameters, similar to Janus TMD nanotubes \cite{bhardwaj2022strain}. Interestingly, the stable symmetry (2H-t/1T-o) found in the present TMH nanotubes is opposite to that in TMD and Janus TMD nanotubes \cite{bhardwaj2021strain, bhardwaj2022strain}. We have also verified the mechanical stability of the flat TMH monolayers by computing their phonon spectra using the density functional perturbation theory (DFPT) feature in the ABINIT code \cite{gonze2002first} (Supplementary Material). These results suggest that the selected nanotubes are also mechanically stable, since the flat monolayer  represents the stressed configuration (i.e, bending stresses) relative to the equilibrium diameter nanotube.

\begin{table}[h!]
\centering
\caption{Equilibrium diameters in nm for the twelve armchair and zigzag Janus TMH nanotubes. The uncertainty in values accounts for the energy differences being smaller than the numerical accuracy  in the calculations, i.e.,  $10^{-6}$ Ha/atom.}
\begin{tabular}{|c|c|c|c|c|c|c|}
\hline
\multirow{3}{*}{M}	&\multicolumn{2}{c|}{MClI} &\multicolumn{2}{c|}{MBrI}	&\multicolumn{2}{c|}{MClBr} \\ \cline{2-7}

&Armchair &Zigzag	&Armchair &Zigzag	&Armchair &Zigzag\\
\hline
Ti & $3.2\pm0.2$ & $3.2\pm0.2$ & $6.2\pm0.4$ & $6.0\pm0.2$ & $7.4\pm0.4$ & $7.8\pm0.4$ \\
\hline
Zr & $4.0\pm0.2$ & $3.8\pm0.2$ & $7.0\pm0.4$ & $6.6\pm0.4$ & $8.6\pm0.8$ & $8.8\pm0.8$ \\
\hline
Hf & $3.8\pm0.2$ & $3.8\pm0.2$ & $6.8\pm0.4$ & $6.8\pm0.4$ & $8.6\pm0.6$ & $8.6\pm0.4$ \\
\hline
Fe & $3.0\pm0.2$ & $2.8\pm0.2$ & $5.8\pm0.2$ & $6.0\pm0.2$ & $5.0\pm0.4$ & $4.8\pm0.4$ \\
\hline
\end{tabular} \label{Table:EqDia}
\end{table}

We calculate the variation in bandgap and effective mass of charge carriers (i.e., electrons and holes) with axial and shear strains using the procedure detailed in Ref.~\cite{sharma2021real}. The axial strain is defined as the  ratio of change in nanotube length to its original length, while the shear strain is defined as the product of nanotube radius and the applied external twist per unit length.  The strain values considered here are commensurate with those found in axial \cite{kaplan2007mechanical, kaplan2006mechanical} and torsion \cite{levi2015nanotube, divon2017torsional, nagapriya2008torsional} experiments,  with the maximum value decided by the mechanical stability of the flat configuration, as determined by DFPT calculations performed  using ABINIT (Supplementary Material).   All numerical parameters in Cyclix-DFT, including those listed above, are chosen such that the bandgap and charge carriers' effective mass are numerically converged to within 0.01 eV and 0.01 a.u., respectively.
    
\section{Results and discussion} \label{Sec:Results}

We now present and discuss the Janus TMH nanotubes' electronic response to axial and torsional deformations, the simulations having been performed using the symmetry-adapted ab initio framework based on Kohn-Sham DFT that has been  described in the previous section.  The more detailed data is available in the Supplementary Material, and wherever possible, the results are compared with those available in literature.

\begin{figure}[h!]
        \centering
        \includegraphics[width=0.95\textwidth]{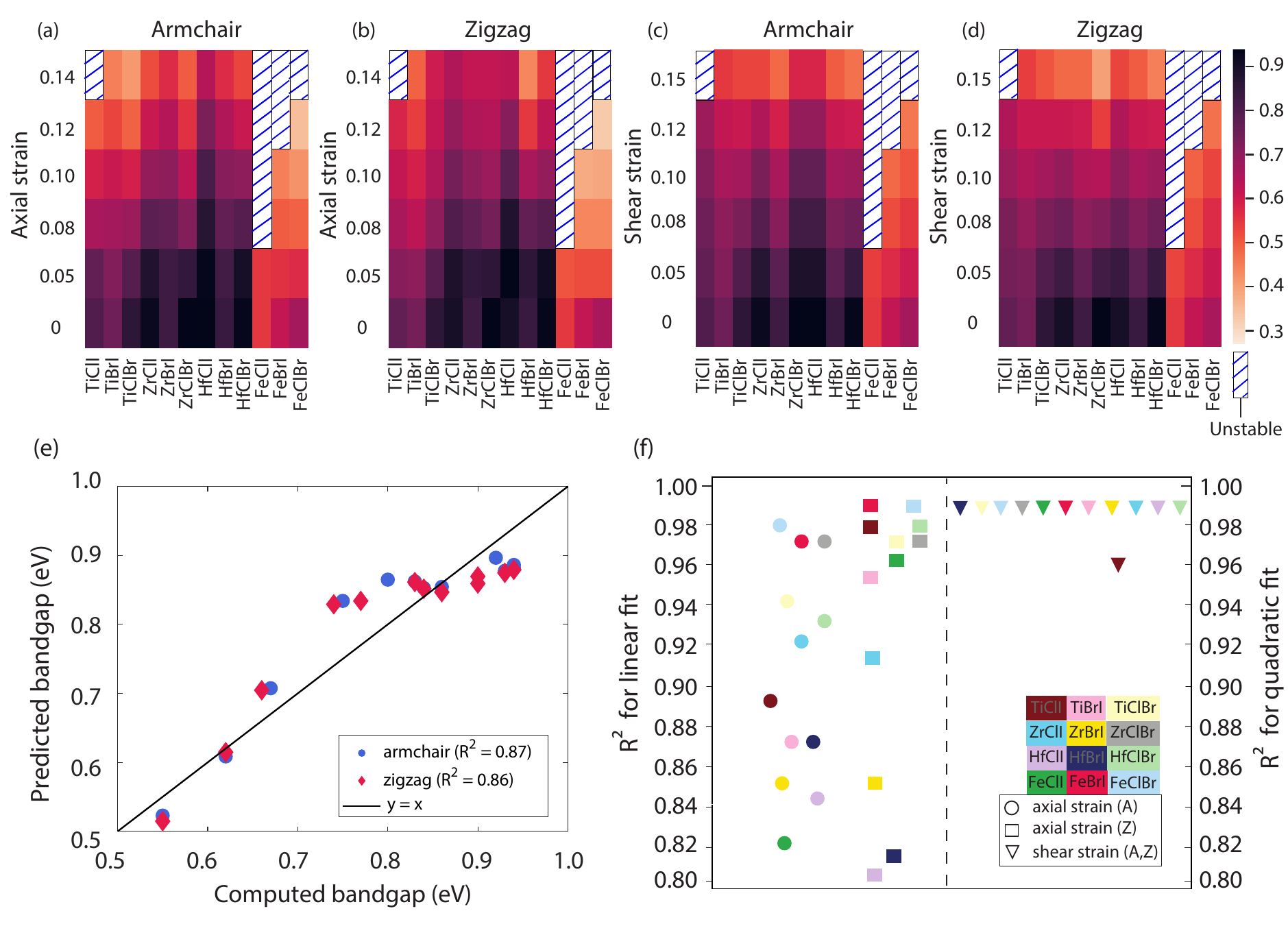}
        \caption{(a), (b): variation of bandgap with axial strains; and (c), (d): variation of bandgap with shear strains for the twelve armchair and zigzag Janus TMH nanotubes. (e): computed bandgap for the undeformed nanotubes vs. that predicted by the linear regression model. (f): coefficient of determination of linear regression ($R^2$) for linear and quadratic fits of bandgap vs. axial and shear strains, respectively.}
      \label{fig:heatbandgap}
    \end{figure}

In figure~\ref{fig:heatbandgap}(a)--(d),  we present the variation in bandgap  with applied axial and shear strains for the Janus TMH nanotubes.  We observe that in the undeformed (i.e., equilibrium) configuration,  all  the nanotubes are semiconducting, with bandgaps ranging from $0.55$ (FeClI) to $0.94$ eV (ZrClBr and HfClI).  For predicting the bandgap variation between the different materials, we develop a regression model using the bond lengths and electronegativity difference between the halogens as features. We find that this simple model is able to capture the bandgaps of the undeformed nanotubes reasonably accurately (figure~\ref{fig:heatbandgap}(e)). Upon the application of axial/torsional deformations, the value of the bandgap reduces monotonically.  In particular,  the bandgap decreases linearly and quadratically with axial and shear strains, respectively, the  average coefficient of determination of linear regression for the linear and quadratic fits being 0.92 and 0.99, respectively (figure~\ref{fig:heatbandgap}(f)).  Similar trends have recently been predicted for the response of transition metal dichalcogenide (TMD) nanotubes \cite{bhardwaj2021strain} and their Janus counterparts \cite{bhardwaj2022strain}. In the case of TMD nanotubes, the linear decrease of bandgap with axial strains has also been predicted in Refs.~\cite{zibouche2014electromechanical, li2014strain, wang2016strain, ghorbani2013electromechanics}. The results indicate that semiconductor to metal transitions are likely to occur when larger strains are considered, however the stability of the nanotubes under such conditions warrants a more careful analysis. The ability to strain engineer the bandgap in Janus TMH nanotubes has potential applications in devices, e.g., mechanical sensors \cite{li2016low, sorkin2014nanoscale, oshima2020geometrical}.

\begin{figure}[h!]
\centering
\includegraphics[width=0.9\textwidth]{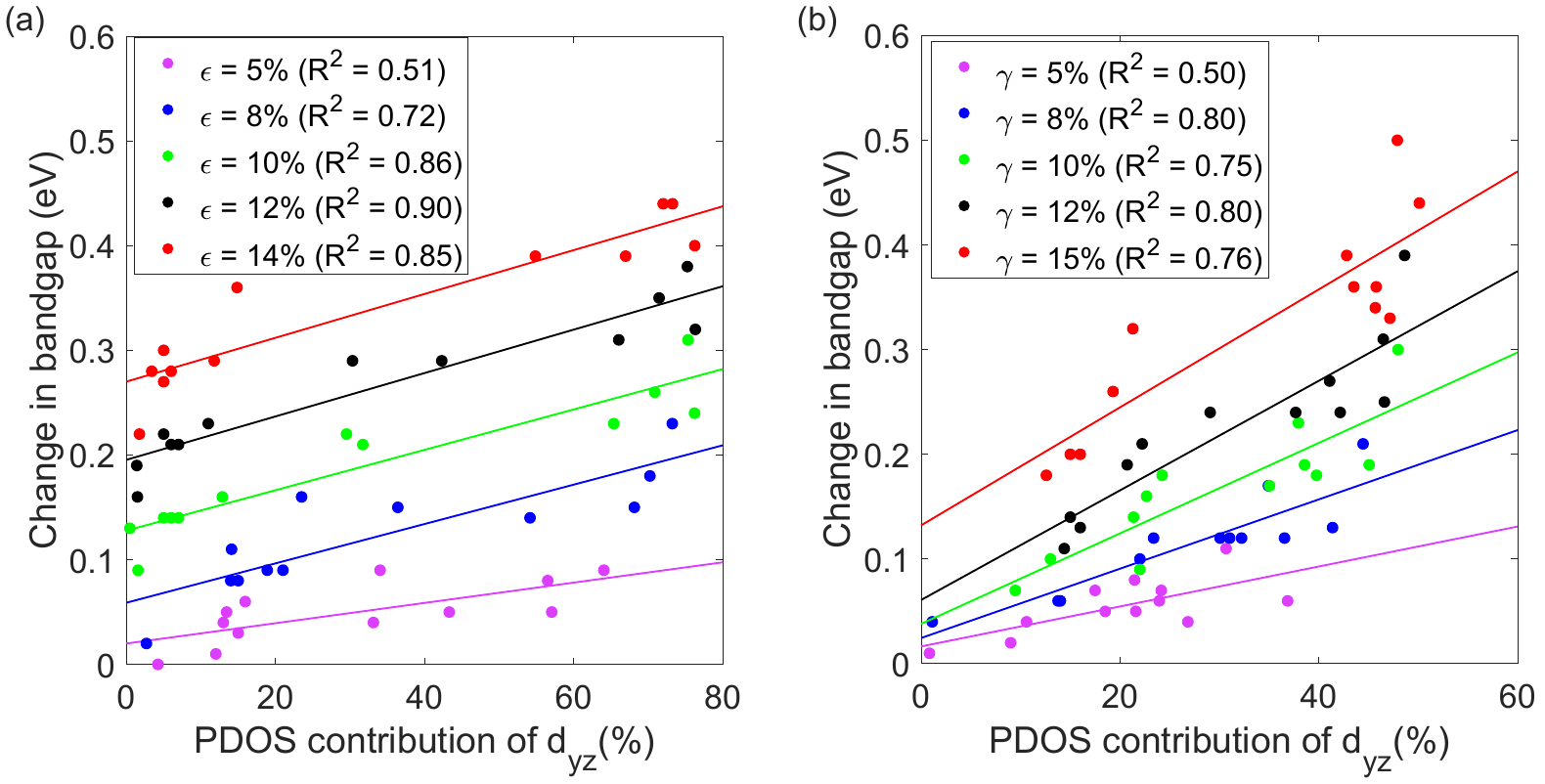}
\caption{The change in bandgap for the twelve armchair Janus TMH nanotubes vs.  the percentage contribution of the transition metal's $d_{yz}$ orbital to the PDOS at the bottom of the conduction band.} \label{fig:pdos}
\end{figure}

To correlate the bandgap variation with changes in the electronic structure, we compute the atomic orbital projected density of states (PDOS) for the Janus TMH nanotubes at all the  configurations studied in this work, i.e., equilibrium and axially/torsionally deformed (Supplementary Material). For armchair nanotubes,  we find  that the transition metal's in-plane $d$ orbital, i.e., $d_{yz}$ orbital, is dominant near the edges of the bandgap --- exception being when the transition metal is iron, for which the $d_{xz}$ orbital is dominant --- making it the focus of study here. In addition, the changes that occur with deformation are more significant at the upper edge of the bandgap, i.e., bottom of the conduction band, therefore the analysis here is restricted to that region. In figure \ref{fig:pdos}, for each of the five axial and shear strains considered, we plot the change in bandgap vs. the contribution of the transition metal's $d_{yz}$ orbital to the PDOS at the bottom of the conduction band. We observe that for a given strain, the change in bandgap is well correlated with the transition metal's $d_{yz}$'s contribution  to the PDOS, with the average coefficient of determination of the linear regression  for axial and shear strains being 0.77 and 0.72, respectively. The importance of $d_{yz}$ in determining the response is to be expected, given its in-plane nature, which means that it is the most likely to undergo significant changes upon the application of axial and torsional deformations.

\begin{figure}[h!]
        \centering
        \includegraphics[width=0.95\textwidth]{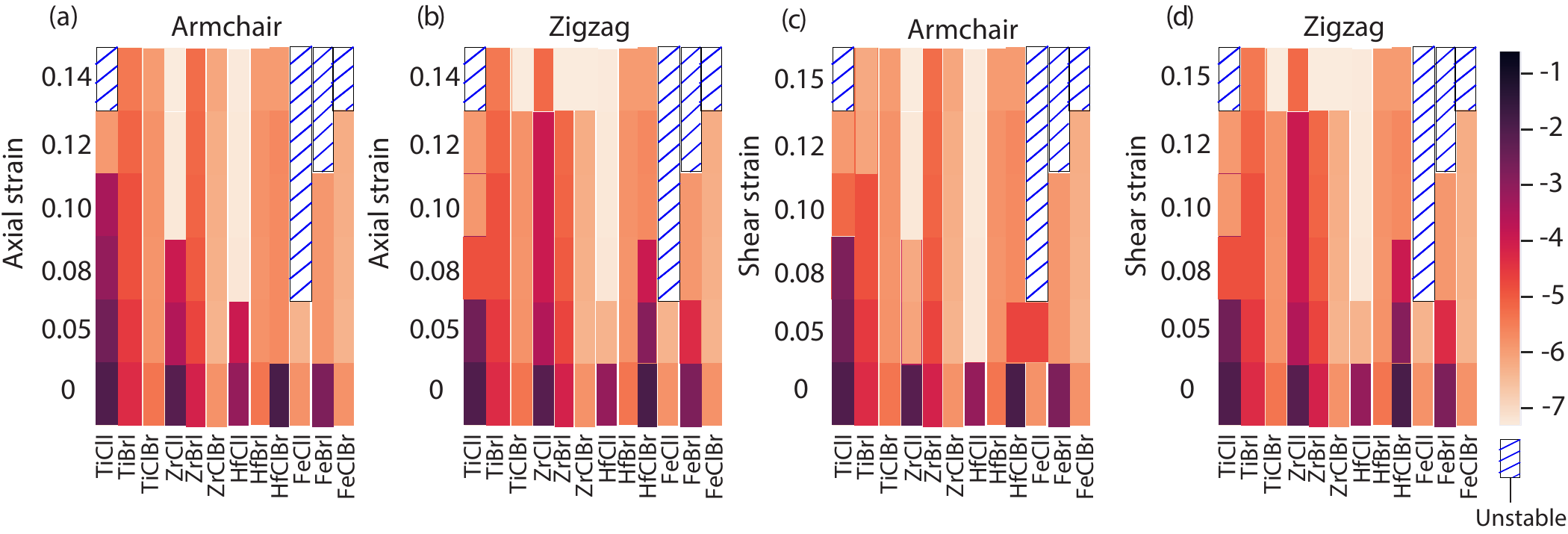}
        \caption{Variation of the difference in  charge carriers' effective mass (holes minus electrons) with axial and shear strains for the twelve armchair and zigzag Janus TMH nanotubes.}
      \label{fig:heatholemass}
    \end{figure}

In figure~\ref{fig:heatholemass}, we present the variation in the difference between the effective mass of holes and electrons (holes minus electrons) with axial and shear strains for the Janus TMH nanotubes. Based on whether this quantity is positive or negative, the material can be classified as a n-type or p-type semiconductor, respectively. In particular, if the effective mass of electron is greater than of the hole, then the mobility of electrons is lower, translating to a lower conduction of electrons, making it a p-type semiconductor, and vice versa \cite{hautier2013identification}. We observe from the figure that all the Janus TMH nanotubes are p-type semiconductors in their undeformed state, differing from TMD and Janus TMD nanotubes, which are generally n-type semiconductors \cite{bhardwaj2021strain, bhardwaj2022strain}. On the application of axial and torsional deformations, there is a continuous increase and decrease in the effective mass of electrons and holes, respectively, similar to that observed for TMD and Janus TMD nanotubes \cite{bhardwaj2021strain, bhardwaj2022strain}.  This hole mobility enhancement \cite{wee2005mobility, mohta2005mobility} makes the Janus TMH nanotubes more dominant p-type semiconductors, which has applications in devices such as MOSFET transistors \cite{mizuno2000electron, maiti1997hole}.

\section{Concluding remarks} \label{Sec:Conclusions}
In this work, we have performed ab initio  Kohn-Sham DFT calculations to study the  electronic response of  Janus TMH nanotubes to mechanical deformations. Specifically, considering twelve armchair and zigzag Janus TMH nanotubes at their equilibrium diameters --- predicted to be stable based on the phonon analysis of flat monolayer counterparts --- we have  determined the variation in bandgap and effective mass of charge carriers with the application of tensile and torsional deformations. We have found that the nanotubes undergo a linear and quadratic decrease in bandgap with tensile and shear strain, respectively. Simultaneously, there is a  continual increase and decrease in the effective mass of electrons and holes, respectively. We have found that for a given strain, the change in bandgap for armchair nanotubes can be correlated with the transition metal's in-plane $d$ orbital's contribution  to the projected density of states at the bottom of the conduction band. 
  
Overall, the current work shows that mechanical deformations represent a powerful tool to control the electronic properties of Janus TMH nanotubes, with applications in devices such as sensors and MOSFET transistors. The study of the optical and thermal response of Janus TMH nanotubes to mechanical deformations presents itself as a worthy  subject for further investigation. The extension of such studies to multi-walled nanotubes is also an interesting topic for research that is currently being pursued by the authors.

\section*{Acknowledgements} 
The authors gratefully acknowledge the support of the U.S. National Science Foundation (CAREER-1553212). This research was supported in part through research cyberinfrastructure resources and services provided by PACE at GT, including the Hive cluster (U.S. National Science Foundation Grant No. MRI-1828187).


\vspace{-1mm}

\begin{thebibliography}{10}

\bibitem{iijima1991helical}
S~Iijima.
\newblock {Helical microtubules of graphitic carbon}.
\newblock {\em Nature}, 354(6348):56--58, 1991.

\bibitem{tenne2003advances}
R~Tenne.
\newblock {Advances in the synthesis of inorganic nanotubes and fullerene-like
  nanoparticles}.
\newblock {\em Angewandte Chemie International Edition}, 42(42):5124--5132,
  2003.

\bibitem{rao2003inorganic}
C~N~R Rao and M~Nath.
\newblock {Inorganic nanotubes}.
\newblock In {\em Advances In Chemistry: A Selection of CNR Rao's Publications
  (1994--2003)}, pages 310--333. World Scientific, 2003.

\bibitem{serra2019overview}
M~Serra, R~Arenal, and R~Tenne.
\newblock {An overview of the recent advances in inorganic nanotubes}.
\newblock {\em Nanoscale}, 11(17):8073--8090, 2019.

\bibitem{ghosh2019symmetry}
S~Ghosh, A~S Banerjee, and P~Suryanarayana.
\newblock {Symmetry-adapted real-space density functional theory for
  cylindrical geometries: Application to large group-IV nanotubes}.
\newblock {\em Physical Review B}, 100(12):125143, 2019.

\bibitem{maiti2003bandgap}
A~Maiti.
\newblock {Bandgap engineering with strain}.
\newblock {\em Nature Materials}, 2(7):440--442, 2003.

\bibitem{zhang2011helical}
D-B Zhang, E~Akatyeva, and T~Dumitric{\u{a}}.
\newblock Helical bn and zno nanotubes with intrinsic twisting: An objective
  molecular dynamics study.
\newblock {\em Physical Review B}, 84(11):115431, 2011.

\bibitem{guo2005systematic}
G~Y Guo and J~C Lin.
\newblock {Systematic ab initio study of the optical properties of BN
  nanotubes}.
\newblock {\em Physical Review B}, 71(16):165402, 2005.

\bibitem{jia2006structure}
J-F Jia, H-S Wu, and H~Jiao.
\newblock {The structure and electronic property of BN nanotube}.
\newblock {\em Physica B: Condensed Matter}, 381(1-2):90--95, 2006.

\bibitem{yang2005electronic}
X~Yang and J~Ni.
\newblock {Electronic properties of single-walled silicon nanotubes compared to
  carbon nanotubes}.
\newblock {\em Physical Review B}, 72(19):195426, 2005.

\bibitem{mintmire1993properties}
J~W Mintmire, D~H Robertson, and C~T White.
\newblock Properties of fullerene nanotubules.
\newblock {\em Journal of Physics and Chemistry of Solids}, 54(12):1835--1840,
  1993.

\bibitem{wang2017band}
C~Wang, X~Fu, Y~Guo, Z~Guo, C~Xia, and Y~Jia.
\newblock {Band gap scaling laws in group IV nanotubes}.
\newblock {\em Nanotechnology}, 28(11):115202, 2017.

\bibitem{ding2002analytical}
J~W Ding, X~H Yan, and J~X Cao.
\newblock Analytical relation of band gaps to both chirality and diameter of
  single-wall carbon nanotubes.
\newblock {\em Physical Review B}, 66(7):073401, 2002.

\bibitem{ouyang2002fundamental}
M~Ouyang, J-L Huang, and C~M Lieber.
\newblock {Fundamental electronic properties and applications of single-walled
  carbon nanotubes}.
\newblock {\em Accounts of Chemical Research}, 35(12):1018--1025, 2002.

\bibitem{wang2019investigation}
H~Wang, N~Ding, T~Jiang, X~Zhao, W~Liu, and F~Za{\"\i}ri.
\newblock {Investigation on mechanical and electronic properties of
  graphene-doped boron nitride nanotubes}.
\newblock {\em Materials Research Express}, 6(11):1150c5, 2019.

\bibitem{parashar2015switching}
V~Parashar, C~P Durand, B~Hao, R~G Amorim, R~Pandey, B~Tiwari, D~Zhang, Y~Liu,
  A-P Li, and Y~K Yap.
\newblock {Switching behaviors of graphene-boron nitride nanotube
  heterojunctions}.
\newblock {\em Scientific Reports}, 5(1):1--6, 2015.

\bibitem{akdim2011bandgap}
B~Akdim and R~Pachter.
\newblock {Bandgap Tuning of a (6, 6) Boron Nitride Nanotube by Analyte
  Physisorption and Application of a Transverse Electric Field: A DFT Study}.
\newblock {\em IEEE Transactions on Nanotechnology}, 10(5):1089--1092, 2011.

\bibitem{tien2005band}
L-G Tien, C-H Tsai, F-Y Li, and M-H Lee.
\newblock {Band-gap modification of defective carbon nanotubes under a
  transverse electric field}.
\newblock {\em Physical Review B}, 72(24):245417, 2005.

\bibitem{chen2004band}
C-W Chen, M-H Lee, and S~J Clark.
\newblock {Band gap modification of single-walled carbon nanotube and boron
  nitride nanotube under a transverse electric field}.
\newblock {\em Nanotechnology}, 15(12):1837, 2004.

\bibitem{fedorov2010tuning}
G~Fedorov, P~Barbara, D~Smirnov, D~Jim{\'e}nez, and Stephan Roche.
\newblock Tuning the band gap of semiconducting carbon nanotube by an axial
  magnetic field.
\newblock {\em Applied Physics Letters}, 96(13):132101, 2010.

\bibitem{li2020unveiling}
Y~Li, W~Liu, H~Xu, H~Chen, H~Ren, J~Shi, W~Du, W~Zhang, Q~Feng, J~Yan, et~al.
\newblock {Unveiling Bandgap Evolution and Carrier Redistribution in Multilayer
  WSe\textsubscript{2}: Enhanced Photon Emission via Heat Engineering}.
\newblock {\em Advanced Optical Materials}, 8(2):1901226, 2020.

\bibitem{nath2003superconducting}
M~Nath, S~Kar, A~K Raychaudhuri, and C~N~R Rao.
\newblock {Superconducting NbSe\textsubscript{2} nanostructures}.
\newblock {\em Chemical Physics Letters}, 368(5-6):690--695, 2003.

\bibitem{tsuneta2003formation}
T~Tsuneta, T~Toshima, K~Inagaki, T~Shibayama, S~Tanda, S~Uji, M~Ahlskog,
  P~Hakonen, and M~Paalanen.
\newblock {Formation of metallic NbSe\textsubscript{2} nanotubes and
  nanofibers}.
\newblock {\em Current Applied Physics}, 3(6):473--476, 2003.

\bibitem{yang2000electronic}
L~Yang and J~Han.
\newblock {Electronic structure of deformed carbon nanotubes}.
\newblock {\em Physical Review Letters}, 85(1):154, 2000.

\bibitem{rochefort1999electrical}
A~Rochefort, P~Avouris, F~Lesage, and D~R Salahub.
\newblock {Electrical and mechanical properties of distorted carbon nanotubes}.
\newblock {\em Physical Review B}, 60(19):13824, 1999.

\bibitem{tombler2000reversible}
T~W Tombler, C~Zhou, L~Alexseyev, J~Kong, H~Dai, L~Liu, C~S Jayanthi, M~Tang,
  and S-Y Wu.
\newblock {Reversible electromechanical characteristics of carbon nanotubes
  underlocal-probe manipulation}.
\newblock {\em Nature}, 405(6788):769--772, 2000.

\bibitem{yang1999band}
L~Yang, M~P Anantram, J~Han, and J~P Lu.
\newblock {Band-gap change of carbon nanotubes: Effect of small uniaxial and
  torsional strain}.
\newblock {\em Physical Review B}, 60(19):13874, 1999.

\bibitem{kim2001electronic}
Y-H Kim, K-J Chang, and S~G Louie.
\newblock {Electronic structure of radially deformed BN and BC\textsubscript{3}
  nanotubes}.
\newblock {\em Physical Review B}, 63(20):205408, 2001.

\bibitem{kinoshita2010electronic}
Y~Kinoshita and N~Ohno.
\newblock {Electronic structures of boron nitride nanotubes subjected to
  tension, torsion, and flattening: A first-principles DFT study}.
\newblock {\em Physical Review B}, 82(8):085433, 2010.

\bibitem{coutinho2009band}
S~S Coutinho, V~Lemos, and S~Guerini.
\newblock {Band-gap tunability of a (6, 0) BN nanotube bundle under pressure:
  Ab initio calculations}.
\newblock {\em Physical Review B}, 80(19):193408, 2009.

\bibitem{ghassemi2012field}
H~M Ghassemi, C~H Lee, Y~K Yap, and R~S Yassar.
\newblock {Field emission and strain engineering of electronic properties in
  boron nitride nanotubes}.
\newblock {\em Nanotechnology}, 23(10):105702, 2012.

\bibitem{qi2012strain}
J~Qi, X~Qian, L~Qi, J~Feng, D~Shi, and J~Li.
\newblock {Strain-engineering of band gaps in piezoelectric boron nitride
  nanoribbons}.
\newblock {\em Nano Letters}, 12(3):1224--1228, 2012.

\bibitem{niquet2012effects}
Y-M Niquet, C~Delerue, and C~Krzeminski.
\newblock {Effects of strain on the carrier mobility in silicon nanowires}.
\newblock {\em Nano Letters}, 12(7):3545--3550, 2012.

\bibitem{kumar2020bending}
S~Kumar and P~Suryanarayana.
\newblock {Bending moduli for forty-four select atomic monolayers from first
  principles}.
\newblock {\em Nanotechnology}, 31(43):43LT01, 2020.

\bibitem{varjovi2021janus}
M~J Varjovi, M~Yagmurcukardes, F~M Peeters, and E~Durgun.
\newblock {Janus two-dimensional transition metal dichalcogenide oxides:
  First-principles investigation of W X O monolayers with X= S, Se, and Te}.
\newblock {\em Physical Review B}, 103(19):195438, 2021.

\bibitem{xiong2018spontaneous}
Q-L Xiong, J~Zhou, J~Zhang, T~Kitamura, and Z-H Li.
\newblock {Spontaneous curling of freestanding Janus monolayer transition-metal
  dichalcogenides}.
\newblock {\em Physical Chemistry Chemical Physics}, 20(32):20988--20995, 2018.

\bibitem{bhardwaj2021elastic}
A~Bhardwaj and P~Suryanarayana.
\newblock {Elastic properties of Janus transition metal dichalcogenide
  nanotubes from first principles}.
\newblock {\em The European Physical Journal B}, 95(1):1--8, 2022.

\bibitem{sreedhara2022nanotubes}
M~B Sreedhara, Y~Miroshnikov, K~Zheng, L~Houben, S~Hettler, R~Arenal, I~Pinkas,
  S~S Sinha, I~E Castelli, and R~Tenne.
\newblock {Nanotubes from Ternary WS\textsubscript{2(1-x)}Se\textsubscript{2x}
  Alloys: Stoichiometry Modulated Tunable Optical Properties}.
\newblock {\em Journal of the American Chemical Society}, 2022.

\bibitem{bhardwaj2022strain}
A~Bhardwaj and P~Suryanarayana.
\newblock {Strain engineering of Janus transition metal dichalcogenide
  nanotubes: an ab initio study}.
\newblock {\em The European Physical Journal B}, 95(3):1--9, 2022.

\bibitem{bhardwaj2021strain}
A~Bhardwaj, A~Sharma, and P~Suryanarayana.
\newblock {Torsional strain engineering of transition metal dichalcogenide
  nanotubes: An ab initio study}.
\newblock {\em Nanotechnology}, 32(47):47LT01, 2021.

\bibitem{yudilevichself}
D~Yudilevich, R~Levi, I~Nevo, R~Tenne, A~Ya’akobovitz, and E~Joselevich.
\newblock {Self-sensing torsional resonators based on inorganic nanotubes}.
\newblock {\em ICME}, pages 1--4, 2018.

\bibitem{levi2015nanotube}
R~Levi, J~Garel, D~Teich, G~Seifert, R~Tenne, and E~Joselevich.
\newblock {Nanotube electromechanics beyond carbon: the case of
  WS\textsubscript{2}}.
\newblock {\em ACS Nano}, 9(12):12224--12232, 2015.

\bibitem{divon2017torsional}
Y~Divon, R~Levi, J~Garel, D~Golberg, R~Tenne, A~Ya’akobovitz, and
  E~Joselevich.
\newblock {Torsional resonators based on inorganic nanotubes}.
\newblock {\em Nano Letters}, 17(1):28--35, 2017.

\bibitem{barua2017nanostructured}
S~Barua, H~S Dutta, R~Gogoi, S~Devi, and R~Khan.
\newblock {Nanostructured MoS\textsubscript{2}-based advanced biosensors: a
  review}.
\newblock {\em ACS Applied Nano Materials}, 1(1):2--25, 2017.

\bibitem{unalan2008zno}
H~E Unalan, Y~Yang, Y~Zhang, P~Hiralal, D~Kuo, S~Dalal, T~Butler, S~N Cha, J~E
  Jang, K~Chremmou, et~al.
\newblock {ZnO Nanowire and WS\textsubscript{2} nanotubes Electronics}.
\newblock {\em IEEE Transactions on Electron Devices}, 55(11):2988--3000, 2008.

\bibitem{zhang2012high}
C~Zhang, S~Wang, L~Yang, Y~Liu, T~Xu, Z~Ning, A~Zak, Z~Zhang, R~Tenne, and
  Q~Chen.
\newblock {High-performance photodetectors for visible and near-infrared lights
  based on individual WS\textsubscript{2} nanotubes}.
\newblock {\em Applied Physics Letters}, 100(24):243101, 2012.

\bibitem{zhang2019enhanced}
Y~J Zhang, T~Ideue, M~Onga, F~Qin, R~Suzuki, A~Zak, R~Tenne, J~H Smet, and
  Y~Iwasa.
\newblock {Enhanced intrinsic photovoltaic effect in tungsten disulfide
  nanotubes}.
\newblock {\em Nature}, 570(7761):349--353, 2019.

\bibitem{bolle2021structural}
F~T B{\"o}lle, A~E~G Mikkelsen, K~S Thygesen, T~Vegge, and I~E Castelli.
\newblock {Structural and chemical mechanisms governing stability of inorganic
  Janus nanotubes}.
\newblock {\em npj Computational Materials}, 7(1):1--8, 2021.

\bibitem{saritas2022piezoelectric}
K~Saritas and S~Ismail-Beigi.
\newblock {Piezoelectric ferromagnetism in two dimensional
  FeCl\textsubscript{2}}.
\newblock {\em arXiv preprint arXiv:2205.00300}, 2022.

\bibitem{li2021room}
R~Li, J~Jiang, W~Mi, and H~Bai.
\newblock {Room temperature spontaneous valley polarization in two-dimensional
  FeClBr monolayer}.
\newblock {\em Nanoscale}, 13(35):14807--14813, 2021.

\bibitem{guo2022substantial}
S-D Guo, J-X Zhu, M-Y Yin, and B-G Liu.
\newblock {Substantial electronic correlation effects on the electronic
  properties in a Janus FeClF monolayer}.
\newblock {\em Physical Review B}, 105(10):104416, 2022.

\bibitem{nath2002nanotubes}
M~Nath and C~N~R Rao.
\newblock {Nanotubes of group 4 metal disulfides}.
\newblock {\em Angewandte Chemie International Edition}, 41(18):3451--3454,
  2002.

\bibitem{bandura2014tis2}
A~V Bandura and R~A Evarestov.
\newblock {TiS\textsubscript{2} and ZrS\textsubscript{2} single-and double-wall
  nanotubes: First-principles study}.
\newblock {\em Journal of Computational Chemistry}, 35(5):395--405, 2014.

\bibitem{sharma2021real}
A~Sharma and P~Suryanarayana.
\newblock {Real-space density functional theory adapted to cyclic and helical
  symmetry: Application to torsional deformation of carbon nanotubes}.
\newblock {\em Physical Review B}, 103(3):035101, 2021.

\bibitem{codony2021transversal}
D~Codony, I~Arias, and P~Suryanarayana.
\newblock Transversal flexoelectric coefficient for nanostructures at finite
  deformations from first principles.
\newblock {\em Physical Review Materials}, 5(3):L030801, 2021.

\bibitem{kumar2021flexoelectricity}
S~Kumar, D~Codony, I~Arias, and P~Suryanarayana.
\newblock {Flexoelectricity in atomic monolayers from first principles}.
\newblock {\em Nanoscale}, 13(3):1600--1607, 2021.

\bibitem{bhardwaj2021torsional}
A~Bhardwaj, A~Sharma, and P~Suryanarayana.
\newblock Torsional moduli of transition metal dichalcogenide nanotubes from
  first principles.
\newblock {\em Nanotechnology}, 32(28):28LT02, 2021.

\bibitem{kumar2022bending}
S~Kumar and P~Suryanarayana.
\newblock {On the bending of rectangular atomic monolayers along different
  directions: an ab initio study}.
\newblock {\em arXiv preprint arXiv:2208.00091}, 2022.

\bibitem{xu2021sparc}
Q~Xu, A~Sharma, B~Comer, H~Huang, E~Chow, A~J Medford, J~E Pask, and
  P~Suryanarayana.
\newblock Sparc: Simulation package for ab-initio real-space calculations.
\newblock {\em SoftwareX}, 15:100709, 2021.

\bibitem{ghosh2017sparc1}
S~Ghosh and P~Suryanarayana.
\newblock {SPARC: Accurate and efficient finite-difference formulation and
  parallel implementation of density functional theory: Isolated clusters}.
\newblock {\em Computer Physics Communications}, 212:189--204, 2017.

\bibitem{ghosh2017sparc2}
S~Ghosh and P~Suryanarayana.
\newblock {SPARC: Accurate and efficient finite-difference formulation and
  parallel implementation of Density Functional Theory: Extended systems}.
\newblock {\em Computer Physics Communications}, 216:109--125, 2017.

\bibitem{banerjee2016cyclic}
A~S Banerjee and P~Suryanarayana.
\newblock {Cyclic density functional theory: A route to the first principles
  simulation of bending in nanostructures}.
\newblock {\em Journal of the Mechanics and Physics of Solids}, 96:605--631,
  2016.

\bibitem{momma2008vesta}
K~Momma and F~Izumi.
\newblock {VESTA: a three-dimensional visualization system for electronic and
  structural analysis}.
\newblock {\em Journal of Applied Crystallography}, 41(3):653--658, 2008.

\bibitem{perdew1996generalized}
J~P Perdew, K~Burke, and M~Ernzerhof.
\newblock {Generalized gradient approximation made simple}.
\newblock {\em Physical Review Letters}, 77(18):3865, 1996.

\bibitem{hamann2013optimized}
D~R Hamann.
\newblock {Optimized norm-conserving Vanderbilt pseudopotentials}.
\newblock {\em Physical Review B}, 88(8):085117, 2013.

\bibitem{spms}
M~F Shojaei, J~E Pask, A~J Medford, and P~Suryanarayana.
\newblock {Soft and transferable pseudopotentials from multi-objective
  optimization}.
\newblock {\em arXiv preprint arXiv:2209.09806}, 2022.

\bibitem{haastrup2018computational}
S~Haastrup, M~Strange, M~Pandey, T~Deilmann, P~S Schmidt, N~F Hinsche, M~N
  Gjerding, D~Torelli, P~M Larsen, A~C Riis-Jensen, et~al.
\newblock {The Computational 2D Materials Database: high-throughput modeling
  and discovery of atomically thin crystals}.
\newblock {\em 2D Materials}, 5(4):042002, 2018.

\bibitem{wu2018tuning}
H-H Wu, Q~Meng, H~Huang, C~T Liu, and X-L Wang.
\newblock {Tuning the indirect--direct band gap transition in the
  MoS\textsubscript{2-x}Se\textsubscript{x} armchair nanotube by diameter
  modulation}.
\newblock {\em Physical Chemistry Chemical Physics}, 20(5):3608--3613, 2018.

\bibitem{tang2018janus}
Z-K Tang, B~Wen, M~Chen, and L-M Liu.
\newblock {Janus MoSSe nanotubes: tunable band gap and excellent optical
  properties for surface photocatalysis}.
\newblock {\em Advanced Theory and Simulations}, 1(10):1800082, 2018.

\bibitem{mikkelsen2021band}
A~E~G Mikkelsen, F~T B{\"o}lle, K~S Thygesen, T~Vegge, and I~E Castelli.
\newblock {Band structure of MoSTe Janus nanotubes}.
\newblock {\em Physical Review Materials}, 5(1):014002, 2021.

\bibitem{luo2019electronic}
Y~F Luo, Y~Pang, M~Tang, Q~Song, and M~Wang.
\newblock {Electronic properties of Janus MoSSe nanotubes}.
\newblock {\em Computational Materials Science}, 156:315--320, 2019.

\bibitem{tao2018band}
L~Tao, Y-Y Zhang, J~Sun, S~Du, and H-J Gao.
\newblock {Band engineering of double-wall Mo-based hybrid nanotubes}.
\newblock {\em Chinese Physics B}, 27(7):076104, 2018.

\bibitem{xie2021theoretical}
S~Xie, H~Jin, Y~Wei, and S~Wei.
\newblock {Theoretical investigation on stability and electronic properties of
  Janus MoSSe nanotubes for optoelectronic applications}.
\newblock {\em Optik}, 227:166105, 2021.

\bibitem{wang2018mechanical}
Y~Z Wang, R~Huang, B~L Gao, G~Hu, F~Liang, and Y~L Ma.
\newblock {Mechanical and strain-tunable electronic properties of Janus MoSSe
  nanotubes}.
\newblock {\em Chalcogenide Letters}, 15(11):535--543, 2018.

\bibitem{gonze2002first}
X~Gonze, J-M Beuken, R~Caracas, F~Detraux, M~Fuchs, G-M Rignanese, L~Sindic,
  M~Verstraete, G~Zerah, F~Jollet, et~al.
\newblock {First-principles computation of material properties: the ABINIT
  software project}.
\newblock {\em Computational Materials Science}, 25(3):478--492, 2002.

\bibitem{kaplan2007mechanical}
I~Kaplan-Ashiri and R~Tenne.
\newblock {Mechanical properties of WS\textsubscript{2} nanotubes}.
\newblock {\em Journal of Cluster Science}, 18(3):549--563, 2007.

\bibitem{kaplan2006mechanical}
I~Kaplan-Ashiri, S~R Cohen, K~Gartsman, V~Ivanovskaya, T~Heine, G~Seifert,
  I~Wiesel, H~D Wagner, and R~Tenne.
\newblock {On the mechanical behavior of WS\textsubscript{2} nanotubes under
  axial tension and compression}.
\newblock {\em Proceedings of the National Academy of Sciences},
  103(3):523--528, 2006.

\bibitem{nagapriya2008torsional}
K~S Nagapriya, O~Goldbart, I~Kaplan-Ashiri, G~Seifert, R~Tenne, and
  E~Joselevich.
\newblock {Torsional stick-slip behavior in WS\textsubscript{2} nanotubes}.
\newblock {\em Physical Review Letters}, 101(19):195501, 2008.

\bibitem{zibouche2014electromechanical}
N~Zibouche, M~Ghorbani-Asl, T~Heine, and A~Kuc.
\newblock {Electromechanical properties of small transition-metal
  dichalcogenide nanotubes}.
\newblock {\em Inorganics}, 2(2):155--167, 2014.

\bibitem{li2014strain}
W~Li, G~Zhang, M~Guo, and Y-W Zhang.
\newblock {Strain-tunable electronic and transport properties of
  MoS\textsubscript{2} nanotubes}.
\newblock {\em Nano Research}, 7(4):518--527, 2014.

\bibitem{wang2016strain}
Y~Z Wang, R~Huang, X~Q Wang, Q~F Zhang, B~L Gao, L~Zhou, and G~Hua.
\newblock {Strain-tunable electronic properties of CrS\textsubscript{2}
  nanotubes}.
\newblock {\em Chalcogenide Letters}, 13(7):301--307, 2016.

\bibitem{ghorbani2013electromechanics}
M~Ghorbani-Asl, N~Zibouche, M~Wahiduzzaman, A~F Oliveira, A~Kuc, and T~Heine.
\newblock {Electromechanics in MoS\textsubscript{2} and WS\textsubscript{2}:
  nanotubes vs. monolayers}.
\newblock {\em Scientific Reports}, 3:2961, 2013.

\bibitem{li2016low}
B~L Li, J~Wang, H~L Zou, S~Garaj, C~T Lim, J~Xie, N~B Li, and D~T Leong.
\newblock {Low-dimensional transition metal dichalcogenide nanostructures based
  sensors}.
\newblock {\em Advanced Functional Materials}, 26(39):7034--7056, 2016.

\bibitem{sorkin2014nanoscale}
V~Sorkin, H~Pan, H~Shi, S~Y Quek, and Y~W Zhang.
\newblock {Nanoscale transition metal dichalcogenides: structures, properties,
  and applications}.
\newblock {\em Critical Reviews in Solid State and Materials Sciences},
  39(5):319--367, 2014.

\bibitem{oshima2020geometrical}
S~Oshima, M~Toyoda, and S~Saito.
\newblock {Geometrical and electronic properties of unstrained and strained
  transition metal dichalcogenide nanotubes}.
\newblock {\em Physical Review Materials}, 4(2):026004, 2020.

\bibitem{hautier2013identification}
G~Hautier, A~Miglio, G~Ceder, G-M Rignanese, and X~Gonze.
\newblock {Identification and design principles of low hole effective mass
  p-type transparent conducting oxides}.
\newblock {\em Nature Communications}, 4(1):1--7, 2013.

\bibitem{wee2005mobility}
C~Wee, S~Maikop, and C~Y Yu.
\newblock {Mobility-enhancement technologies}.
\newblock {\em IEEE Circuits and Devices Magazine}, 21(3):21--36, 2005.

\bibitem{mohta2005mobility}
N~Mohta and Scott~E Thompson.
\newblock {Mobility enhancement}.
\newblock {\em IEEE Circuits and Devices Magazine}, 21(5):18--23, 2005.

\bibitem{mizuno2000electron}
T~Mizuno, S~Takagi, N~Sugiyama, H~Satake, A~Kurobe, and A~Toriumi.
\newblock {Electron and hole mobility enhancement in strained-Si MOSFET's on
  SiGe-on-insulator substrates fabricated by SIMOX technology}.
\newblock {\em IEEE Electron Device Letters}, 21(5):230--232, 2000.

\bibitem{maiti1997hole}
C~K Maiti, L~K Bera, S~S Dey, D~K Nayak, and N~B Chakrabarti.
\newblock {Hole mobility enhancement in strained-Si p-MOSFETs under high
  vertical field}.
\newblock {\em Solid-State Electronics}, 41(12):1863--1869, 1997.

\end{thebibliography}
\end{document}